%% file: main.tex
\documentclass[runningheads]{llncs}
\usepackage[T1]{fontenc}
\usepackage{graphicx}
\usepackage{romannum}
\usepackage{booktabs}
\usepackage{lscape}
\usepackage{tablefootnote}
\usepackage[table]{xcolor}
\begin{document}
\title{Enhancing LLM-Based Feedback: Insights from Intelligent Tutoring Systems and the Learning Sciences}
\titlerunning{Enhancing LLM-Based Feedback}

\author{John Stamper\inst{1}\orcidID{0000-0002-2291-1468} \and
Ruiwei Xiao\inst{1}\orcidID{0000-0002-6461-7611} \and
Xinying Hou\inst{2}\orcidID{0000-0002-1182-5839}}

\authorrunning{J. Stamper et al.}
%
\institute{
Carnegie Mellon University, Pittsburgh, PA 15213, USA\\
\email{\{jstamper,ruiweix\}@cs.cmu.edu}\and 
University of Michigan, Ann Arbor, MI 48109, USA\\
\email{xyhou@umich.edu}
}

\maketitle              

\begin{abstract}
    \input{sections/00_abstract}

\end{abstract}
\section{Introduction}
  \input{sections/01_Introduction}
  
\section{The Development of ITS Feedback Generation}
  \input{sections/03_development}
  
\section{Implications for LLM-based feedback}
  \input{sections/04_learning}

\section{Conclusion}
  \input{sections/06_conclusion}

\bibliographystyle{splncs04}
\bibliography{citations} 
\end{document}

%% file: sections/00_abstract.tex
The field of Artificial Intelligence in Education (AIED) focuses on the intersection of technology, education, and psychology, placing a strong emphasis on supporting learners' needs with compassion and understanding. The growing prominence of Large Language Models (LLMs) has led to the development of scalable solutions within educational settings, including generating different types of feedback in Intelligent Tutoring Systems. However, the approach to utilizing these models often involves directly formulating prompts to solicit specific information, lacking a solid theoretical foundation for prompt construction and empirical assessments of their impact on learning. This work advocates careful and caring AIED research by going through previous research on feedback generation in ITS, with emphasis on the theoretical frameworks they utilized and the efficacy of the corresponding design in empirical evaluations, and then suggesting opportunities to apply these evidence-based principles to the design, experiment, and evaluation phases of LLM-based feedback generation. The main contributions of this paper include: an avocation of applying more cautious, theoretically grounded methods in feedback generation in the era of generative AI; and practical suggestions on theory and evidence-based feedback design for LLM-powered ITS. 
\keywords{Intelligent Tutoring System (ITS) \and Large Language Models (LLMs) \and Generative AI (GenAI) \and Hint \and Formative Feedback.}

%% file: sections/01_introduction.tex
While the term Generative AI (GenAI) has become synonymous with LLMs and exploded with the release of ChatGPT in 2022 \cite{baidoo2023education}, it is important to note that GenAI has been a part of the AIED community for many years. A long line of work has been seen in areas where LLMs have shown particular usefulness such as content generation \cite{di2005natural}, including generating formative feedback \cite{stamper2010enhancing}. The rush to apply LLMs to these areas to improve AIED certainly represents a good opportunity to help in education, but the community should not discount the decades of work that has been done in these areas before the current interest in the latest LLMs. In particular, the AIED community has decades of research on the proper way to implement hints and feedback, and while LLMs seem to be helpful in different phases of feedback generation, the approaches used should be carefully designed and evaluated through theories and empirical evidence. 

This work sets forth a balanced discourse on the integration of LLMs with learning science work, focusing on leveraging both previous insights on feedback design in intelligent tutoring systems (ITSs) and contemporary advancements in generative AI. It outlines a strategic blueprint for infusing LLM-based feedback by building on previous contributions around ITSs and the learning sciences, aiming to refine feedback generation processes and achieve more effective learning results. This work underscores the significance of adhering to established educational frameworks and validates the potential of LLMs to improve feedback components in educational systems, thus offering a pathway toward more effective and responsible future educational technologies.

%% file: sections/03_development.tex
In this section, we synthesized existing research on ITS feedback generation based on how they were generated. Three primary methods were identified for generating feedback: the expert-created learner model, the data-driven learner model, and the use of large language models. The first method involves experts manually inputting models of learners' potential behaviors or constraints of the problem. The second one compiles data from learners' interactions with similar problems to automatically build a model of learner behavior. The third method leverages LLMs, focusing on supplying the appropriate context and requirements, striving to generate more adaptive feedback with less human labor.





\subsection{Feedback generated from expert-generated learner model}

To generate feedback for students, traditional ITSs heavily rely on experts' input on learner modeling, with an emphasis on student problem-solving states. There are two main lines of learner modeling methods in these intelligent tutors: the production rules model, which originated from Anderson's ACT-R theory \cite{anderson1996act}, and the constraint-based model (CBM), which is based on Ohlsson's theory of learning from performance errors \cite{ohlsson1996learning}. Specifically, for a given problem, a production rules model uses a set of if-then rules or example solutions \cite{aleven2009new} generated by experts to model the knowledge and possible decision-making process to solve this problem. A CBM is composed of a set of expert-written constraints that should not be broken by any potential correct solutions. When generating feedback, the learner's solution will be compared to the rule-based model or constraint-based model, and the point the student gets astray from the right track will be included in the delivered feedback. 

The effectiveness of expert-generated models for both approaches has been repeatedly proven by evidence from theories and classroom studies. For example, feedback in CTAT tutors \cite{aleven2006cognitive} is generated from production rules. Until 2016, there were 18 CTAT tutors distributed in real educational settings, used by approximately 44,000 students. One of the CTAT tutors, Andes Physics Tutor \cite{vanlehn2005andes} with its immediate feedback, helping students achieve significantly better learning in five years' repeated measurements. CBM has also proven to be extremely effective and efficient on highly structured procedural tasks and open-ended tasks such as programming. For instance, the feedback generated by CBM in SQL-Tutor led to significantly higher learning outcomes in 4 studies during 1998-2000 \cite{mitrovic2001constraint}. The learning curve analysis further grounded the experiment result with sound psychological foundations aligning with the smooth learning curve criterion \cite{martin2005using}.

\subsection{Feedback generated from data-driven learner model}
Regardless of their effectiveness, building expert models can be extremely taxing, making them hard to scale up. Furthermore, these methods have inherent limitations: experts may overlook common mistakes made by students, and both strategies often yield less-optimal results in ill-defined domains \cite{fournier2010learning}. Therefore, researchers have started to seek scalable solutions by applying data-driven methods to aggregate previous students' solutions and construct the learner's model. The automated feedback generation nature of the learner model positions the data-driven feedback approach as an initial application of generative AI in creating feedback. 

The earliest work of generative AI on feedback generation can be traced back to the DIAG system \cite{di2005natural}. The system utilized the NLP model to aggregate system messages in various structures and found the feedback aggregated by functions led to higher learning gain in the classroom study. Another stream of data-driven works built on the production rule approach automated the construction of the learner's cognitive model during problem-solving. The foundation work of the data-driven cognitive model, regardless of its non-generative nature, can be found in 1997 \cite{conati1997line}, where researchers initially applied Bayesian Networks to students' problem-solving data for plan recognition and action prediction. This approach was further developed by the Hint Factory \cite{stamper2008hint}, which employed Markov decision processes to analyze Logic Proof Intelligent Tutor submission data \cite{stamper2006automating}, generating production rules and comparing student submissions to these rules to tailor hints. This method of production-rule-based feedback generation garnered attention across ITS research in more disciplines \cite{fossati2015data},\cite{fournier2010learning},\cite{price2016generating}. The subsequent classroom studies and learning factor analyses \cite{cen2006learning} confirmed this approach's efficacy and versatility across various domains, demonstrating its potential with minimal data requirements \cite{rivers2017data}.

\subsection{Feedback generated from LLMs}
Some major shortcomings of the data-driven approach are: 1) it relies on the quantity and quality of the training data, and 2) the feedback in these systems is in fixed templates with limited adaptation. The recent prevalence of LLMs provides opportunities to advance the field of feedback generation by generating adaptive, human-like feedback without training data. For the feedback generation pipeline, most LLM-based work takes the student's current state together with certain prompts asking for feedback as input, treats the LLMs as a black box to process the prompt, and directly uses the output as the personalized feedback. 

Indeed, LLMs enhance the scalability of adaptive feedback across various domains \cite{mcnichols2023algebra,nguyen2023evaluating,kazemitabaar2024codeaid,lee2023smartphone,schmucker2023ruffle}, and help in bypassing the expert model or cognitive model building process from scratch. However, there remains a significant amount of work overlooked by many researchers in the pedagogical design of feedback and the evaluation of its impact on learning. Few works are backed by learning sciences principles such as learning-by-teaching \cite{schmucker2023ruffle} and self-reflection \cite{kumar_quickta_nodate}, while others did not elaborate learning design considerations in their design rationale. Moreover, most of the evaluations on the LLM-based feedback systems only reported classroom usage data, and there is a lack of theoretical support or evidence on learning for these emerging systems. To guide better LLM-based feedback design and evaluation, in the next section, we highlight theoretical and empirical evidence from previous ITS and learning sciences work, hence suggesting implications for LLM-based feedback accordingly. 

%% file: sections/04_learning.tex
\begin{figure}
\centering
\includegraphics[width=\textwidth]{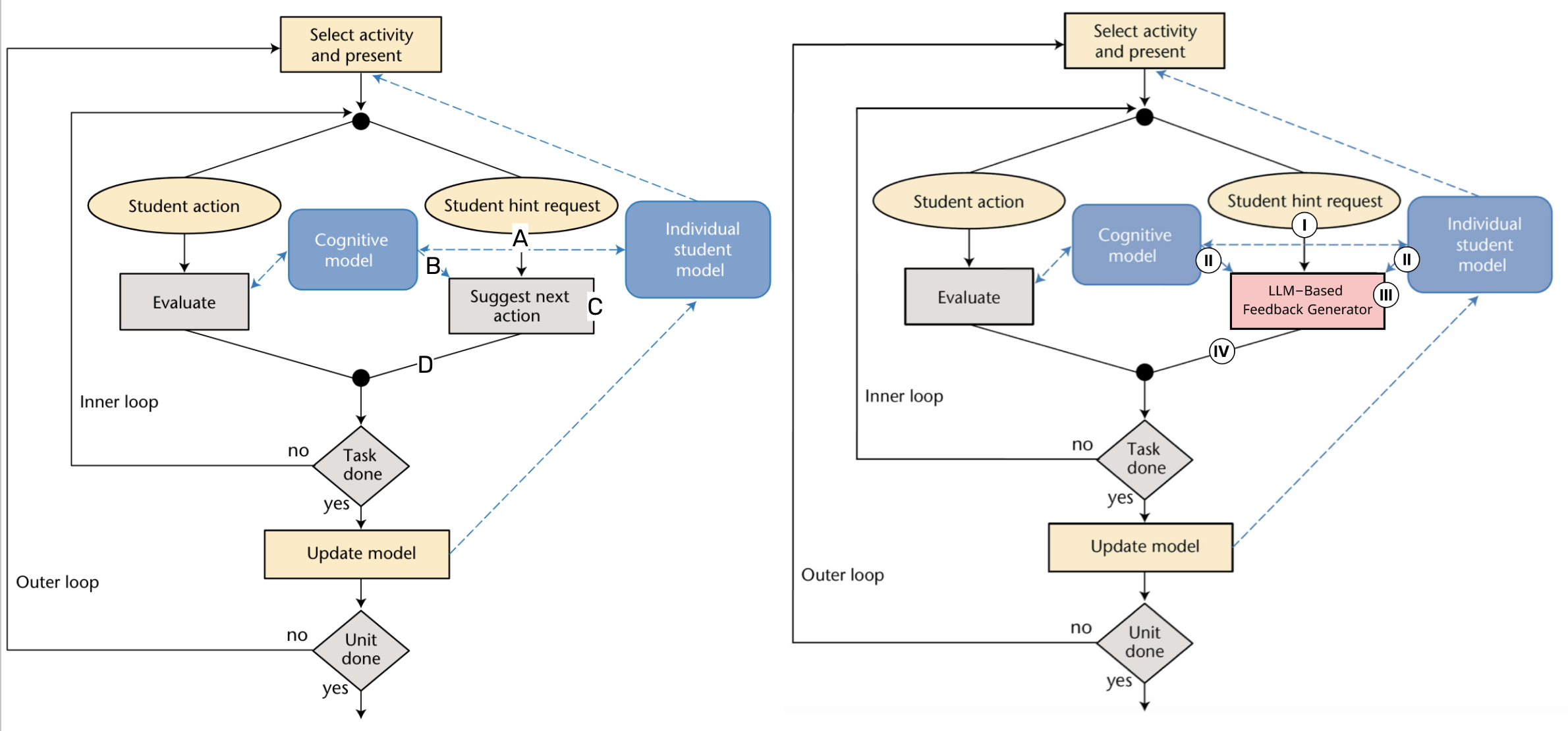}
\caption{Key functions of Intelligent Tutoring Systems from Koedinger et al. \cite{Koedinger2013NewPF} (left) and Opportunities to Optimize LLM-based Feedback in ITS (right, adapted from \cite{Koedinger2013NewPF}): \Romannum{1} - Trigger to deliver the feedback; \Romannum{2} - Information needed to communicate with LLMs as input; \Romannum{3} - Content requested in the generated LLM feedback output; \Romannum{4} - Modality for delivering feedback to students.} \label{potentials_in_its_workflow}
\vspace{-6mm}
\end{figure}

Prior research in ITSs has built strong groundwork for LLM-based feedback generation. For instance, the left figure in Figure \ref{potentials_in_its_workflow} illustrates the key functions of a traditional intelligent tutoring system, showing how feedback is triggered, generated, and delivered in an ITS \cite{Koedinger2013NewPF}. However, methods proven effective in traditional ITS have not yet been fully leveraged in LLM-based feedback research. Additionally, technological advancements in GenAI now enable scaling approaches previously limited by technological constraints. 

This section discusses recommendations grounded in prior ITS research and new scalable opportunities from GenAI advancements in the feedback design, generation, delivery, and evaluation phases. The right figure in Figure \ref{potentials_in_its_workflow} shows how new opportunities in LLM and GenAI could fit into and extend the traditional key functions in ITS (the left one in Figure \ref{potentials_in_its_workflow}), particularly around the feedback.
\subsection{Trigger to deliver the feedback}
Feedback in ITSs is often designed to be triggered either by the system or initiated by the students \cite{aleven2009new,mitrovic2013effect}. One special type of feedback is suggestions or supportive materials to help students move forward. It is common to provide such feedback based on students' request actions \cite{hou2022using,hou2023understanding,mclaren2022instructional,xiao2024exploring,rivers2017data}, as shown in A of Figure \ref{potentials_in_its_workflow}-left. However, in some other work, concerning that novices might lack the metacognitive skills to identify when and what to ask for help, researchers believed that the model-guided feedback delivery could scaffold learners better. Therefore, Help Tutor was designed to provide metacognitive feedback automatically after the system identifies learners’ misbehaviors \cite{aleven2006toward}. While researchers found that having this did not lead to higher learning gains, the concerns about abusing feedback can be more serious when LLM is widely applied in educational settings. On the one hand, students nowadays can access commercial LLM products easily for timely formative feedback on their progress or even direct answers, so they are more likely to abuse them by asking for feedback too frequently. On the other hand, to avoid over-reliance on these LLM tools, some self-regulated learners might avoid asking for formative feedback, even when they are struggling or unsure about their answers. With those two new phenomena in mind, future research needs to find a balanced timing and optimized trigger to deliver feedback and establish a more effective learning experience (\Romannum{1} in Figure \ref{potentials_in_its_workflow}-right).

\vspace{-3mm}

\subsection{Information needed to communicate with LLM as input}
Prompt engineering is an important step when using LLMs to generate feedback in the existing learning systems. Prompt engineering refers to the practice of creating and optimizing prompts to communicate with large language models effectively. There are guidelines about how to generate effective prompts for LLMs. For example, the CLEAR framework comprises five key principles to achieve prompts for more effective AI-generated content creation \cite{lo2023clear}. Besides these general guidelines, the required information in prompts can vary based on the specific roles AI is expected to take. When AI takes on the role of providing feedback to students' answer inputs \cite{nguyen2023evaluating} \cite{lu2021integrating}, information resources like the grading rubric, the student answers, and the task description, are commonly included in the prompt.  While current implications have already integrated such basic related information to generate prompts, one future direction could involve incorporating various types of information into prompt engineering. For example, many traditional ITSs represent students' current understanding of the subject as a model and adapt instructions to students' needs based on it \cite{vanlehn2013student} (B in Figure \ref{potentials_in_its_workflow}-left). Such student models play an important role in understanding and identifying students' needs before offering appropriate adaptation \cite{chrysafiadi2013student}. Therefore, to enhance the quality of LLM outputs in meeting individual student's needs, different layers of information from a student model, such as the estimations of the student’s knowledge level, cognitive state, strengths or weaknesses, affective state, and meta-cognitive skills, can be integrated when conducting prompt engineering in the context of education \cite{chrysafiadi2013student} (\Romannum{2} in Figure \ref{potentials_in_its_workflow}-right).

\subsection{Content requested in the generated LLM feedback output}
One type of ITS feedback output is a suggestion for the next action (C in Figure \ref{potentials_in_its_workflow}-left). When integrating LLM into learning systems, such output can also be optimized using learning science theories (\Romannum{3} in Figure \ref{potentials_in_its_workflow}-right). Particularly, the prevalent form of feedback in recent LLM-based learning systems is a high-level natural language explanation without revealing the answer. In certain systems, persistent student inquiries for the answer lead the feedback to progressively reveal more comprehensive hints, culminating in responses that closely resemble complete, bottom-out solutions. Although such feedback can assist students in solving their problems, its effectiveness in promoting long-term learning is uncertain. Should this feedback fail to facilitate learning, the LLM-based Intelligent Tutoring System (ITS) might merely become an ineffective variant or, in the best-case scenario, a duplicate of ChatGPT. To better trigger learning, this section aims to encourage future feedback implementation guided by learning sciences frameworks such as Bloom's Taxonomy \cite{krathwohl2002revision} and Knowledge-Learning-Instruction (KLI) framework \cite{koedinger2012knowledge}, and we employ the KLI framework as a guiding example to illustrate how to select feedback content based on the knowledge type to enhance learner outcomes. The KLI framework allocates learning processes into three types: memory and fluency, induction and refinement, understanding, and sense-making, and lists seven instructional principles that are effectively robust within each category, all of which are backed by substantial experimental validation. 

For enhancing memory and fluency, instructional principles such as spacing, testing, and optimized scheduling are pivotal. They focus on the timing and frequency of practice repetitions. Feedback mechanisms like flashcards, visual cues \cite{lee2023smartphone}, or multiple-choice questions, inspired by the testing principle, offer repeated practice opportunities at short intervals, reinforcing learning. While spacing and optimization extend beyond feedback's scope, their implementation in ITS is beneficial to effective instructional strategies in memory and fluency processes.

For tasks involving induction and refinement processes, such as solving programming or math problems, faded worked examples are effective in demonstrating the desired process with the appropriate level of assistance needed by fading the demonstration to ensure the scaffolding is within Vygotsky’s Zone of Proximal Development \cite{malik2017revisiting}, tailored to the learner's existing knowledge and areas for growth. The implementation of faded worked examples into intelligent tutors also resulted in higher learning efficiency and a deeper conceptual understanding of the problems in existing ITS works \cite{schwonke2007can}.

For understanding and sense-making processes, self-explanation questions are one of the most effective forms of feedback that elicit active learning and deeper thinking \cite{hausmann2007explaining}. Previous ITS works also support its effects on consolidating learners' understanding of fractions \cite{rau2009intelligent}. 

\subsection{Modality for delivering feedback to students}
Due to technology limitations, traditional ITS systems mainly deliver feedback to students as text (D in Figure \ref{potentials_in_its_workflow}-left). Recent advancements in GenAI techniques provide opportunities to expand the spectrum of feedback modalities (\Romannum{4} in Figure \ref{potentials_in_its_workflow}-right) from text to images, audios, videos, and combinations of these modalities. As the main focus of this paper is LLM-based feedback, we look into how GenAI techniques could deliver such feedback to students in diverse modalities. We first referred to classic multimedia learning principles \cite{clark2023learning} and chose related principles that provide recommendations on how feedback might be delivered. Then, we provided examples of recent GenAI techniques and corresponding commercial applications that could be used to achieve such feedback modalities. Finally, for each principle-based feedback modality, we provided one example use scenario. We summarized these in Table \ref{modality}. As new GenAI technologies are continuously growing, we believe it is important for future work to look at both the principle and technology sides to decide what feedback modalities should and could be provided in different learning contexts.

\subsection{Evaluate the quality of the generated feedback}
Existing studies has tested the scope of LLMs' feedback generation capabilities with data-driven system evaluation (whether the system can perform well on a comprehensive testing dataset) \cite{hellas2023exploring} and expert evaluation (given an evaluation matrix, the expert(s) would rate the performance of each feedback) \cite{nguyen2023evaluating,roest2024next,xiao2024exploring} on mainly precision and coverage\cite{phung2023generating,phung2024automating}. Moreover, it is important to recognize the contributions of GenAI-generated feedback as it expands the array of metrics for assessing feedback quality beyond the traditional ITS evaluation scope, with new metrics like appropriateness, conciseness, and comprehensiveness. Compared to traditional feedback generation approaches, LLMs' advantages in natural language tasks enable the generation of personalized feedback with an encouraging tone \cite{roest2024next,xiao2024exploring}. Taking a step further from examining the tone, studies have also assessed the capability of GenAI to play specific roles (e.g., instructor \cite{tack2022ai}, student\cite{schmucker2023ruffle}) in the context of feedback generation. These initial assessments illuminate the potential to scale both the implementation and evaluation of specific agents and their behaviors that are crafted according to learning sciences principles, such as facilitating a growth mindset as an instructor or enhancing the dynamics of a collaborative learning environment as a helpful peer.

In addition to system and expert evaluation, classroom deployment allows researchers to gather feedback from real learners in authentic educational settings. Under this context, student usage log data \cite{qi2023conversational,kazemitabaar2024codeaid,kumar_quickta_nodate,kazemitabaar2023novices} and student self-reported survey data \cite{kazemitabaar2024codeaid,kumar_quickta_nodate,pankiewicz2024navigating} are two commonly gathered data types for LLM-generated material evaluation. With few existing LLM-generated feedback works applying these approaches, the next steps for evaluating this feedback could involve: 1) designing controlled experiments with pre-post tests to assess learning outcomes \cite{vanlehn2005andes,price2017hint,alhazmi2020interactive}, and 2) employing learning analytics on log data to explore feedback's effectiveness on help-seeking \cite{xiao2024exploring} and learning \cite{pankiewicz2024navigating} with greater detail and nuances. For instance, the impacts of feedback on learning outcomes can be elucidated by performing a learning curve analysis to compare the slopes, which represent varying learning rates under different feedback conditions. \cite{rivers2016learning}. Finally, ethical concerns such as biases \cite{wei2024uncovering} and hallucination \cite{chen2023hallucination} associated with LLMs could jeopardize an effective and equitable learning environment. Therefore, it is crucial to incorporate relevant metrics into the evaluation of feedback quality and mitigate these issues with automated approaches or human intervention.




%% file: sections/06_conclusion.tex
To fully utilize the potential of generative AI in the educational context, it is essential to approach its integration with pedagogical design. This paper synthesized the progression of AIED focusing on feedback generation, emphasized the importance of grounding the current LLM-based feedback generation in theoretical frameworks and evidence-based approaches to prompt its learning effectiveness, and suggested corresponding evidence-grounded implications through four feedback generation stages. We aim to evoke the awareness of AIED researchers and practitioners on the legacy of pre-LLM feedback generation efforts, and offer a toolkit as a foundational resource for designing pedagogical LLM-based feedback generation systems that foster the advancement of the AIED field in the era of generative AI.
\begin{landscape}
\begin{table}
\caption{Example feedback modalities supported by Generative AI techniques}
\begin{minipage}{\textwidth}
\rowcolors{2}{gray!30!}{gray!10!}
\begin{tabular}{p{2.5cm}p{4cm}p{6cm}p{6cm}}
\toprule
\begin{tabular}[c]{@{}l@{}}Principle \\ (Selected from \cite{clark2023learning})\end{tabular}   & \multicolumn{1}{c}{Feedback modality} & {Example GenAI Techniques Involved (Example GenAI Applications)}                                           & \multicolumn{1}{c}{Example Feedback Scenario}                                                                                                                  \\ \midrule
Multimedia                         & Images                                & \begin{tabular}[c]{@{}l@{}}Text-to-image (OpenAI DALL·E \cite{openai2023dalle2}; \\ Stable Diffusion \footnote{https://stability.ai/stable-image}; Runway \footnote{https://runwayml.com/ai-tools/text-to-image/})\end{tabular} & Content-related images as visual cue for vocabulary memorization.                                                                                               \\
Personalization                    & Text                                  & \begin{tabular}[c]{@{}l@{}}Text-to-text\\ (OpenAI GPT \footnote{https://openai.com/gpt-4}; Anthropic Claude \footnote{https://www.anthropic.com/news/claude-3-family};\\  Meta Llama\footnote{https://llama.meta.com/})\end{tabular}                                                                   & Text feedback in more informal and conversational styles.                                                                                                       \\
Embodiment                         & Human-like agent                      & \begin{tabular}[c]{@{}l@{}}Text-to-image; \\ Text-to-video \\  (Runway\footnote{https://runwayml.com/ai-tools/gen-2-text-to-video/}; OpenAI Sora \cite{openai_sora})\\Text-to-speech \\ (Deepgram \footnote{https://deepgram.com/}; OpenAI TTS models \footnote{https://platform.openai.com/docs/models/tts}; \\ WellSaid Labs \footnote{https://wellsaidlabs.com/features/api/})\end{tabular} & Virtual teaching assistant avatar talking about the next-step hints.                                                                                             \\
Modality                           & Audio                                 & Text-to-speech                                                                                                                                                      & Add audio feedback as a new option to existing systems.                                                                                                        \\
Segmentation                       & Segmented Text                        & Text-to-text                                                                                                                                                        & Apply LLMs to segment one feedback into multi-levels of feedback. \\
 \bottomrule

\end{tabular}
\label{modality}
\end{minipage}
\end{table}
\end{landscape}

\begin{landscape}
\begin{table}
\rowcolors{2}{gray!30!}{gray!10!}
\begin{tabular}{p{2.5cm}p{4cm}p{5cm}p{6.5cm}}
\toprule
\begin{tabular}[c]{@{}l@{}}Principle \\ (Selected from \cite{clark2023learning})\end{tabular}   & \multicolumn{1}{c}{Feedback modality} & \begin{tabular}[c]{@{}l@{}} Example GenAI Techniques Involved \\ (Example GenAI Applications)\end{tabular}                                               & \multicolumn{1}{c}{Example Feedback Scenario}                                                                                                                  \\ \midrule
Redundancy                         & Audio, Image, Video                   & \begin{tabular}[c]{@{}l@{}}Text-to-speech; Text-to-image; \\ Text-to-video \end{tabular}   & When both audio and visual feedback is generated, allow users to turn off the text on screen.    \\
Temporal Contiguity                & Audio, Image, Video                   & \begin{tabular}[c]{@{}l@{}}Text-to-speech; Text-to-image; \\ Text-to-video \end{tabular}                                                                          & When both audio and visual feedback are generated, present them simultaneously. \\
Spatial Contiguity                 & Text, Video                           & Text-to-text; Text-to-video                                                                                                & Place feedback close to the part that it elaborates on.                                                                                                        \\
Coherence                          & Text, Image                           & Text-to-text; Text-to-image                                                                                                & Evaluate whether existing content are on-topic or not.                                                                                                          \\
Signaling                          & Text                                  & Text-to-text                                                                                                                                                        & Highlight the errors on the screen.                                                                                                                                 \\ \bottomrule
\end{tabular}
\end{table}
\end{landscape}

%% file: main.bbl
\begin{thebibliography}{10}
\providecommand{\url}[1]{\texttt{#1}}
\providecommand{\urlprefix}{URL }
\providecommand{\doi}[1]{https://doi.org/#1}

\bibitem{aleven2006toward}
Aleven, V., Mclaren, B., Roll, I., Koedinger, K.: Toward meta-cognitive tutoring: A model of help seeking with a cognitive tutor. International Journal of Artificial Intelligence in Education  \textbf{16}(2),  101--128 (2006)

\bibitem{aleven2006cognitive}
Aleven, V., McLaren, B.M., Sewall, J., Koedinger, K.R.: The cognitive tutor authoring tools (ctat): Preliminary evaluation of efficiency gains. In: Intelligent Tutoring Systems: 8th International Conference, ITS 2006, Jhongli, Taiwan, June 26-30, 2006. Proceedings 8. pp. 61--70. Springer (2006)

\bibitem{aleven2009new}
Aleven, V., Mclaren, B.M., Sewall, J., Koedinger, K.R.: A new paradigm for intelligent tutoring systems: Example-tracing tutors. International Journal of Artificial Intelligence in Education  \textbf{19}(2),  105--154 (2009)

\bibitem{alhazmi2020interactive}
Alhazmi, S., Thevathayan, C., Hamilton, M.: Interactive pedagogical agents for learning sequence diagrams. In: Artificial Intelligence in Education: 21st International Conference, AIED 2020, Ifrane, Morocco, July 6--10, 2020, Proceedings, Part II 21. pp. 10--14. Springer (2020)

\bibitem{anderson1996act}
Anderson, J.R.: Act: A simple theory of complex cognition. American psychologist  \textbf{51}(4), ~355 (1996)

\bibitem{baidoo2023education}
Baidoo-Anu, D., Ansah, L.O.: Education in the era of generative artificial intelligence (ai): Understanding the potential benefits of chatgpt in promoting teaching and learning. Journal of AI  \textbf{7}(1),  52--62 (2023)

\bibitem{cen2006learning}
Cen, H., Koedinger, K., Junker, B.: Learning factors analysis--a general method for cognitive model evaluation and improvement. In: International conference on intelligent tutoring systems. pp. 164--175. Springer (2006)

\bibitem{chen2023hallucination}
Chen, Y., Fu, Q., Yuan, Y., Wen, Z., Fan, G., Liu, D., Zhang, D., Li, Z., Xiao, Y.: Hallucination detection: Robustly discerning reliable answers in large language models. In: Proceedings of the 32nd ACM International Conference on Information and Knowledge Management. pp. 245--255 (2023)

\bibitem{chrysafiadi2013student}
Chrysafiadi, K., Virvou, M.: Student modeling approaches: A literature review for the last decade. Expert Systems with Applications  \textbf{40}(11),  4715--4729 (2013)

\bibitem{clark2023learning}
Clark, R.C., Mayer, R.E.: E-learning and the science of instruction: Proven guidelines for consumers and designers of multimedia learning. john Wiley \& sons (2023)

\bibitem{conati1997line}
Conati, C., Gertner, A.S., VanLehn, K., Druzdzel, M.J.: On-line student modeling for coached problem solving using bayesian networks. In: User Modeling: Proceedings of the Sixth International Conference UM97 Chia Laguna, Sardinia, Italy June 2--5 1997. pp. 231--242. Springer (1997)

\bibitem{di2005natural}
Di~Eugenio, B., Fossati, D., Yu, D., Haller, S.M., Glass, M.: Natural language generation for intelligent tutoring systems: a case study. In: AIED. pp. 217--224 (2005)

\bibitem{fossati2015data}
Fossati, D., Di~Eugenio, B., Ohlsson, S., Brown, C., Chen, L.: Data driven automatic feedback generation in the ilist intelligent tutoring system. Technology, Instruction, Cognition and Learning  \textbf{10}(1),  5--26 (2015)

\bibitem{fournier2010learning}
Fournier-Viger, P., Nkambou, R., Nguifo, E.M.: Learning procedural knowledge from user solutions to ill-defined tasks in a simulated robotic manipulator. Romero, et al.(eds.) Handbook of Educational Data Mining pp. 451--465 (2010)

\bibitem{hausmann2007explaining}
Hausmann, R.G., VanLehn, K.: Explaining self-explaining: A contrast between content and generation. Frontiers in Artificial Intelligence and Applications  \textbf{158}, ~417 (2007)

\bibitem{hellas2023exploring}
Hellas, A., Leinonen, J., Sarsa, S., Koutcheme, C., Kujanp{\"a}{\"a}, L., Sorva, J.: Exploring the responses of large language models to beginner programmers’ help requests. In: Proceedings of the 2023 ACM Conference on International Computing Education Research-Volume 1. pp. 93--105 (2023)

\bibitem{hou2022using}
Hou, X., Ericson, B.J., Wang, X.: Using adaptive parsons problems to scaffold write-code problems. In: Proceedings of the 2022 ACM Conference on International Computing Education Research-Volume 1. pp. 15--26 (2022)

\bibitem{hou2023understanding}
Hou, X., Ericson, B.J., Wang, X.: Understanding the effects of using parsons problems to scaffold code writing for students with varying cs self-efficacy levels. In: Proceedings of the 23rd Koli Calling International Conference on Computing Education Research. pp. 1--12 (2023)

\bibitem{kazemitabaar2023novices}
Kazemitabaar, M., Hou, X., Henley, A., Ericson, B.J., Weintrop, D., Grossman, T.: How novices use llm-based code generators to solve cs1 coding tasks in a self-paced learning environment. In: Proceedings of the 23rd Koli Calling International Conference on Computing Education Research. pp. 1--12 (2023)

\bibitem{kazemitabaar2024codeaid}
Kazemitabaar, M., Ye, R., Wang, X., Henley, A.Z., Denny, P., Craig, M., Grossman, T.: Codeaid: Evaluating a classroom deployment of an llm-based programming assistant that balances student and educator needs. arXiv preprint arXiv:2401.11314  (2024)

\bibitem{Koedinger2013NewPF}
Koedinger, K., Brunskill, E., Baker, R., Mclaughlin, E., Stamper, J.C.: New potentials for data-driven intelligent tutoring system development and optimization. AI Mag.  \textbf{34},  27--41 (2013), \url{https://api.semanticscholar.org/CorpusID:13189100}

\bibitem{koedinger2012knowledge}
Koedinger, K.R., Corbett, A.T., Perfetti, C.: The knowledge-learning-instruction framework: Bridging the science-practice chasm to enhance robust student learning. Cognitive science  \textbf{36}(5),  757--798 (2012)

\bibitem{krathwohl2002revision}
Krathwohl, D.R.: A revision of bloom's taxonomy: An overview. Theory into practice  \textbf{41}(4),  212--218 (2002)

\bibitem{kumar_quickta_nodate}
Kumar, H., Musabirov, I., Williams, J.J., Liut, M.: Quickta: Exploring the design space of using large language models to provide support to students. In: Learning Analytics and Knowledge Conference. Learning Analytics and Knowledge Conference 2023 (LAK’23), ACM, Arlington, Texas (2023)

\bibitem{lee2023smartphone}
Lee, J., Lan, A.: Smartphone: Exploring keyword mnemonic with auto-generated verbal and visual cues. In: International Conference on Artificial Intelligence in Education. pp. 16--27. Springer (2023)

\bibitem{lo2023clear}
Lo, L.S.: The clear path: A framework for enhancing information literacy through prompt engineering. The Journal of Academic Librarianship  \textbf{49}(4),  102720 (2023)

\bibitem{lu2021integrating}
Lu, C., Cutumisu, M.: Integrating deep learning into an automated feedback generation system for automated essay scoring. International Educational Data Mining Society  (2021)

\bibitem{malik2017revisiting}
Malik, S.A.: Revisiting and re-representing scaffolding: The two gradient model. Cogent Education  \textbf{4}(1),  1331533 (2017)

\bibitem{martin2005using}
Martin, B., Koedinger, K.R., Mitrovic, A., Mathan, S.: On using learning curves to evaluate its. In: AIED. pp. 419--426 (2005)

\bibitem{mclaren2022instructional}
McLaren, B.M., Richey, J.E., Nguyen, H., Hou, X.: How instructional context can impact learning with educational technology: Lessons from a study with a digital learning game. Computers \& education  \textbf{178},  104366 (2022)

\bibitem{mcnichols2023algebra}
McNichols, H., Zhang, M., Lan, A.: Algebra error classification with large language models. In: International Conference on Artificial Intelligence in Education. pp. 365--376. Springer (2023)

\bibitem{mitrovic2001constraint}
Mitrovic, A., Mayo, M., Suraweera, P., Martin, B.: Constraint-based tutors: a success story. In: Engineering of Intelligent Systems: 14th International Conference on Industrial and Engineering Applications of Artificial Intelligence and Expert Systems, IEA/AIE 2001 Budapest, Hungary, June 4--7, 2001 Proceedings 14. pp. 931--940. Springer (2001)

\bibitem{mitrovic2013effect}
Mitrovic, A., Ohlsson, S., Barrow, D.K.: The effect of positive feedback in a constraint-based intelligent tutoring system. Computers \& Education  \textbf{60}(1),  264--272 (2013)

\bibitem{nguyen2023evaluating}
Nguyen, H.A., Stec, H., Hou, X., Di, S., McLaren, B.M.: Evaluating chatgpt’s decimal skills and feedback generation in a digital learning game. In: European Conference on Technology Enhanced Learning. pp. 278--293. Springer (2023)

\bibitem{ohlsson1996learning}
Ohlsson, S.: Learning from performance errors. Psychological review  \textbf{103}(2), ~241 (1996)

\bibitem{openai2023dalle2}
{OpenAI}: Dall·e 2. \url{https://openai.com/dall-e-2/} (2023), accessed: 2024-03-12

\bibitem{openai_sora}
{OpenAI}: {Sora}. \url{https://openai.com} (2024), accessed: 2024-03-12

\bibitem{pankiewicz2024navigating}
Pankiewicz, M., Baker, R.S.: Navigating compiler errors with ai assistance--a study of gpt hints in an introductory programming course. arXiv preprint arXiv:2403.12737  (2024)

\bibitem{phung2023generating}
Phung, T., Cambronero, J., Gulwani, S., Kohn, T., Majumdar, R., Singla, A., Soares, G.: Generating high-precision feedback for programming syntax errors using large language models. arXiv preprint arXiv:2302.04662  (2023)

\bibitem{phung2024automating}
Phung, T., P{\u{a}}durean, V.A., Singh, A., Brooks, C., Cambronero, J., Gulwani, S., Singla, A., Soares, G.: Automating human tutor-style programming feedback: Leveraging gpt-4 tutor model for hint generation and gpt-3.5 student model for hint validation. In: Proceedings of the 14th Learning Analytics and Knowledge Conference. pp. 12--23 (2024)

\bibitem{price2016generating}
Price, T.W., Dong, Y., Barnes, T.: Generating data-driven hints for open-ended programming. International Educational Data Mining Society  (2016)

\bibitem{price2017hint}
Price, T.W., Zhi, R., Barnes, T.: Hint generation under uncertainty: The effect of hint quality on help-seeking behavior. In: Artificial Intelligence in Education: 18th International Conference, AIED 2017, Wuhan, China, June 28--July 1, 2017, Proceedings 18. pp. 311--322. Springer (2017)

\bibitem{qi2023conversational}
Qi, J.Z.P.L., Hartmann, B., Norouzi, J.D.N.: Conversational programming with llm-powered interactive support in an introductory computer science course. NeurIPS’23 Workshop on Generative AI for Education (GAIED)  (2023)

\bibitem{rau2009intelligent}
Rau, M.A., Aleven, V., Rummel, N.: Intelligent tutoring systems with multiple representations and self-explanation prompts support learning of fractions. In: AIED. pp. 441--448 (2009)

\bibitem{rivers2016learning}
Rivers, K., Harpstead, E., Koedinger, K.R.: Learning curve analysis for programming: Which concepts do students struggle with? In: ICER. vol.~16, pp. 143--151. ACM (2016)

\bibitem{rivers2017data}
Rivers, K., Koedinger, K.R.: Data-driven hint generation in vast solution spaces: a self-improving python programming tutor. International Journal of Artificial Intelligence in Education  \textbf{27},  37--64 (2017)

\bibitem{roest2024next}
Roest, L., Keuning, H., Jeuring, J.: Next-step hint generation for introductory programming using large language models. In: Proceedings of the 26th Australasian Computing Education Conference. pp. 144--153 (2024)

\bibitem{schmucker2023ruffle}
Schmucker, R., Xia, M., Azaria, A., Mitchell, T.: Ruffle\&riley: Towards the automated induction of conversational tutoring systems. arXiv preprint arXiv:2310.01420  (2023)

\bibitem{schwonke2007can}
Schwonke, R., Wittwer, J., Aleven, V., Salden, R., Krieg, C., Renkl, A.: Can tutored problem solving benefit from faded worked-out examples. In: European Cognitive Science Conference. pp. 23--27 (2007)

\bibitem{stamper2006automating}
Stamper, J.: Automating the generation of production rules for intelligent tutoring systems. In: Proc. 9th Int. Conf. Interact. Comput. Aided Learn (2006)

\bibitem{stamper2010enhancing}
Stamper, J., Barnes, T., Croy, M.: Enhancing the automatic generation of hints with expert seeding. In: Intelligent Tutoring Systems: 10th International Conference, ITS 2010, Pittsburgh, PA, USA, June 14-18, 2010, Proceedings, Part II 10. pp. 31--40. Springer (2010)

\bibitem{stamper2008hint}
Stamper, J., Barnes, T., Lehmann, L., Croy, M.: The hint factory: Automatic generation of contextualized help for existing computer aided instruction. In: Proceedings of the 9th International Conference on Intelligent Tutoring Systems Young Researchers Track. pp. 71--78 (2008)

\bibitem{tack2022ai}
Tack, A., Piech, C.: The ai teacher test: Measuring the pedagogical ability of blender and gpt-3 in educational dialogues. arXiv preprint arXiv:2205.07540  (2022)

\bibitem{vanlehn2013student}
VanLehn, K.: Student modeling. Foundations of intelligent tutoring systems pp. 55--78 (2013)

\bibitem{vanlehn2005andes}
VanLehn, K., Lynch, C.F., Schulze, K.G., Shapiro, J.A., Shelby, R., Taylor, L., Treacy, D., Weinstein, A., Wintersgill, M.: The andes physics tutoring system: Five years of evaluations. In: AIED. pp. 678--685 (2005)

\bibitem{wei2024uncovering}
Wei, Y., Carvalho, P.F., Stamper, J.: Uncovering name-based biases in large language models through simulated trust game. arXiv preprint arXiv:2404.14682  (2024)

\bibitem{xiao2024exploring}
Xiao, R., Hou, X., Stamper, J.: Exploring how multiple levels of gpt-generated programming hints support or disappoint novices. arXiv preprint arXiv:2404.02213  (2024)

\end{thebibliography}
